# Star formation efficiency across large-scale galactic environments

Laya Ghodsi,[1,2]★ Allison W. S. Man[1]★ Darko Donevski,[3,4] Romeel Davé,[5,6,7] Seunghwan Lim,[8,9] Christopher C. Lovell[10,11] and Desika Narayanan[12,13]

[1]*Department of Physics & Astronomy, University of British Columbia, 6224 Agricultural Road, Vancouver, BC V6T 1Z1, Canada*
[2]*ESO, Karl Schwarzschild strasse 2, D-85748 Garching, Germany*
[3]*National Centre for Nuclear Research, Pasteura 7, PL-02-093 Warsaw, Poland*
[4]*SISSA, Via Bonomea 265, I-34136 Trieste, Italy*
[5]*Institute for Astronomy, University of Edinburgh, Royal Observatory, Blackford Hill, Edinburgh EH9 3HJ, UK*
[6]*Department of Physics, University of the Western Cape, Bellville, Cape Town 7535, South Africa*
[7]*South African Astronomical Observatories, Observatory, Cape Town 7925, South Africa*
[8]*Kavli Institute for Cosmology, University of Cambridge, Madingley Road, Cambridge CB3 0HA, UK*
[9]*Cavendish Laboratory, University of Cambridge, 19 JJ Thomson Avenue, Cambridge CB3 0HE, UK*
[10]*Institute of Cosmology and Gravitation, University of Portsmouth, Burnaby Road, Portsmouth PO1 3FX, UK*
[11]*Astronomy Centre, University of Sussex, Falmer, Brighton BN1 9QH, UK*
[12]*Department of Astronomy, University of Florida, 211 Bryant Space Sciences Center, Gainesville, FL 32611 USA*
[13]*Cosmic Dawn Center (DAWN), Niels Bohr Institute, University of Copenhagen, Jagtvej 128, DK-2200 København N, Denmark*

## ABSTRACT

Environmental effects on the formation and evolution of galaxies have been one of the leading questions in galaxy studies during the past few decades. In this work, we investigate the relationship between the star formation activity of galaxies and their environmental matter density using the cosmological hydrodynamic simulation SIMBA. The galactic star formation activity indicators that we explore include the star formation efficiency (SFE), specific star formation rate (sSFR), and molecular hydrogen mass fraction ($f^*_{H_2}$), and the environment is considered as the large-scale environmental matter density, calculated based on the stellar mass of nearby galaxies on a 1 $h^{-1}$ Mpc grid using the cloud in cell method. Our sample includes galaxies with $9 < \log \frac{M_*}{M_\odot}$ at $0 < z < 4$, divided into three stellar mass bins to disentangle the effects of stellar mass and environment on the star formation activity of galaxies. For low- to intermediate-mass galaxies at low redshifts ($z < 1.5$), we find that the star formation efficiency of those in high-density regions are ∼0.3 dex lower than those in low-density regions. However, there is no significant environmental dependence of the star formation efficiency for massive galaxies over all our redshift range, and low- to intermediate-mass galaxies at high redshifts ($z > 1.5$). We present a scaling relation for the depletion time of cold molecular hydrogen ($t_{\rm depl} = 1/{\rm SFE}$) as a function of galaxy parameters including environmental density. Our findings provide a framework for quantifying the environmental effects on the star formation activities of galaxies as a function of stellar mass and redshift. The most significant environmental dependence is seen at later cosmic times ($z < 1.5$) and towards lower stellar masses ($9 < \log \frac{M_*}{M_\odot} < 10$). Future large galaxy surveys can use this framework to look for the environmental dependence of the star formation activity and examine our predictions.

## 1 INTRODUCTION

It is well established that the environment of galaxies affect their properties including colour (Balogh et al. 2004; Bamford et al. 2009), morphology (Dressler 1980; Goto et al. 2003; Skibba et al. 2009), and star formation rate (SFR; Kauffmann et al. 2004; Peng et al. 2010; Woo et al. 2013; Old et al. 2020). These works show that in the low-redshift Universe ($z < 1$), galaxies in dense regions tend to be redder, more elliptical, and less star-forming. In contrast to the low-redshift Universe, there is no consensus in the literature about the relationship between SFR and the environment at high redshift ($z > 1$). Some studies claim that at high redshift, galaxies in dense regions have higher SFRs than galaxies in low-density regions, on average (Elbaz et al. 2007; Cooper et al. 2008; Santos et al. 2014, 2015; Shimakawa et al. 2018). This suggested change in the relationship between SFR and the environment at roughly $z \sim 1$ is called the reversal of the SFR–density relation. This trend partially arises from the virialization process of galaxy clusters. The overdense regions in the low-redshift Universe are mostly virialized galaxy clusters subject to physical processes including starvation and ram-pressure stripping that can reduce the star formation activity of the member galaxies. Whereas, the overdense regions in the higher redshift Universe are mostly non-virialized protoclusters and have large intergalactic gas reservoirs to fuel the star formation in galaxies. However, other observational and theoretical studies find no evidence for a relationship between SFR and the environment of galaxies (Scoville et al. 2013; Darvish et al. 2016; Duivenvoorden et al. 2016; Lovell et al. 2021) or find the same relationship as the low-redshift Universe (Patel et al. 2009). Observational studies on this topic suffer from selection biases due to observational limits that restrict the sample redshift and stellar mass ranges and different methods to define and measure environments and galaxy properties (Muldrew et al. 2012; Muldrew, Hatch & Cooke 2015; Lovell, Thomas & Wilkins 2018). Exploring this research question using hydrodynamic simulations helps in understanding the background physics of galaxy formation and evolution producing the observed trends (Yajima et al. 2022). Although simulations also suffer from known limitations, they have been able to predict many observational trends. Regarding the galactic SFR–density relation, Tonnesen & Cen (2014) and Hwang, Shin & Song (2019) have been able to reproduce the reversal of SFR–density relation using hydrodynamic simulations up to redshifts $z = 1$ and $z = 2$, respectively.

★ E-mail: layaghodsi@phas.ubc.ca (LG); aman@phas.ubc.ca (AWSM)

Cold molecular gas, mainly consisting of molecular hydrogen, is the star formation fuel in galaxies. Studying the effects of the environment on the molecular hydrogen content is thus essential to dissect the role of the environment on the star formation activity of galaxies. Observational studies find different trends for the role of the environment on galactic molecular gas content. At low-redshift, a few studies find no correlation between the environment and the molecular gas content of galaxies (Kenney & Young 1989; Lavezzi & Dickey 1998; Koyama et al. 2017). However, some works report that molecular gas content increases with the environmental density (Mok et al. 2016) while others report that molecular gas content decreases with environmental density (Fumagalli et al. 2009; Scott et al. 2013; Boselli et al. 2014). Molecular gas observations for high-redshift galaxies ($z > 1.5$) are currently quite limited in sample sizes and affected by selection biases. Nevertheless, some works at this redshift range find higher gas fractions for cluster galaxies than the field galaxies (Noble et al. 2017; Hayashi et al. 2018), while other works find no environmental dependence for gas fraction (Lee et al. 2017; Tadaki et al. 2019).

Since SFR and molecular hydrogen content, two tracers for galactic star formation activity, are correlated, it is useful to define another parameter that relates these two parameters to trace the variations of SFR and molecular hydrogen mass at the same time. Star formation efficiency (SFE) is typically defined as the SFR per molecular hydrogen mass (SFE $\equiv$ SFR/$M_{H_2}$; Young et al. 1996; Boselli et al. 2001). Similarly, depletion time ($t_{\rm depl} = 1/{\rm SFE} = M_{H_2}/{\rm SFR}$) is the time-scale during which the galaxy converts all of its molecular hydrogen into new stars at the current star formation rate. A fall in SFR could be due to a lower molecular hydrogen mass supply, and/or less efficient star formation. It is thus important to monitor these three galactic star formation activity parameters together to gain a comprehensive understanding of galaxy evolution (Scoville et al. 2017; Lu et al. 2022).

Previous studies have tried to quantify the SFE and gas depletion time as functions of other galaxy properties. Scoville et al. (2017) quantified the gas depletion time as a function of redshift, star formation rate, and stellar mass. Similarly, Tacconi et al. (2018) introduced a scaling relation for the depletion time as a function of redshift, star formation rate, stellar mass, and the radius of galaxies.

These works, however, lack the environmental effects on galactic star formation activity. Darvish et al. (2018b) include a number of environmental indicators to the Scoville et al. (2017) scaling relation, but they do not find an environmental dependence of the gas depletion time in their sample of galaxies with $\log \frac{M_*}{M_\odot} > 10$ at $0.5 < z < 3.5$. It is worth mentioning that most of the known scaling relations are for star-forming galaxies in the field, without explicit measurements of the environment they reside in (Tacconi et al. 2018; Liu et al. 2019). Systematic investigations of the star formation efficiency of galaxies versus the environment remain sparse to date.

To derive more complete knowledge of galaxy gas properties, we must also investigate whether galaxy large-scale environment regulates the link between the gas and star formation. The environment of galaxies is defined in a variety of ways based on the available data set and the goal of different studies. On large scales, the matter content in the Universe is distributed in a web-like structure, called the cosmic web. This structure consists of dense massive nodes, long filaments connecting the nodes, vast thin walls, and huge low-density regions called voids. Some works define the environment of galaxies as the cosmic web component in which the galaxy resides (Hahn et al. 2007; Moorman et al. 2016; Xu et al. 2020). However, the complexity of this structure and the wide range of different physical definitions for the components of the cosmic web makes it complicated to study the environmental effects on galaxies using this environment tracer. In addition, membership in a galaxy cluster or field is another definition for galactic environment (Vulcani et al. 2013; Zavala et al. 2019; Lemaux et al. 2020). The result of studies using this method is a strong function of the definition of cluster/field galaxies for which there is not a consensus in the community. At high redshift ($z > 2$), this environment measure is even more tricky given the lack of massive virialized galaxy clusters. Moreover, this dichotomy does not cover the full dynamical range of different galactic environments. The environmental density of galaxies is another measure of the environment. As summarized in Muldrew et al. (2012) and Etherington & Thomas (2015), two commonly used methods in the literature to measure environmental density include: (i) the density within a volume around the galaxy determined by the distance between the galaxy and its $N^{\rm th}$ nearest neighbour, where $N$ typically ranges from 5 to 10 (Bamford et al. 2009; Ellison et al. 2010; Wu 2020) and (ii) the density within a cell, sphere, or cylinder around the galaxy determined by a fixed aperture (Berrier et al. 2011; Smolčić et al. 2017). These methods are computationally fast to perform. The results of these methods might be functions of $N$ for the first method and the fixed aperture size for the second method. The nearest neighbour-based methods probe the internal trends of massive haloes better, while the fixed aperture-based method is a better probe for scales larger than individual haloes (Muldrew et al. 2012). Variations of the two methods based on the environmental density could use two/three-dimensional measures and number/luminosity/mass density which makes the comparison of different studies on this topic even more complicated and tricky. Moreover, the peculiar motion of galaxies along the line of sight, leading to the Kaiser effect (Finger of God), makes it complicated to measure the exact densities (Kaiser 1987). The reader can refer to Haas, Schaye & Jeeson-Daniel (2012), Muldrew et al. (2012), and Darvish et al. (2015) for a more detailed discussion about different environment indicators. In this study, we use the environmental density as an environment indicator calculated considering the stellar mass of neighbour galaxies on a $1\,h^{-1}$ Mpc grid (see Section 3.1).

Although observational capabilities have significantly advanced recently, their ability to detect faint galaxies is still limited by the sensitivity of the instrument. Hence, observed galaxy catalogues, specifically at high redshift, are not complete and suffer from selection biases. This limitation, in addition to various approaches used to measure galactic properties from observations, makes it difficult to capture bias-free trends in galaxy properties. Using simulations is a fruitful method to model galaxies theoretically, avoid observational selection biases, and explore the physical processes that drive the observed trends. In this work, using a set of simulated galaxies from the SIMBA hydrodynamic cosmological simulation (Davé et al. 2019), we aim to shed new light on a question: Does the environment of galaxies affect their star formation activity? We investigate the galactic star formation activity parameters, including SFE, sSFR, molecular hydrogen mass fraction, and depletion time, as functions of redshift, stellar mass, and environmental density. The data set used spans a relatively wide redshift range ($0 < z < 4$), wide stellar mass range ($9 < \log \frac{M_*}{M_\odot} < 13$), wide environmental density range ($-4 < \log [1 + \delta^*_{\rm gal}] < 2$), and a large number of galaxies ($\sim 330\,000$ galaxies). The environmental density is calculated using the stellar mass density of neighbour galaxies on a $1\,h^{-1}$ Mpc grid, calculated using the Cloud in Cell (CIC) method.

This paper is structured as follows: Section 2 presents the properties of the SIMBA simulation and the data we use. In Section 3, we explain our methods to calculate the environmental density and error propagation. In Section 4, we present our results. Section 5 discusses the results and possible interpretations. Finally, Section 6 summarizes the main findings of the work.

## 2 THE SIMBA SIMULATION

In this work, we study the environmental dependence of the star formation activity of galaxies in SIMBA, a state-of-the-art cosmological hydrodynamic simulation. In this section, we briefly explain the main methods and physical prescriptions used to derive the galaxy properties in this simulation.

SIMBA (Davé et al. 2019) is a suite of galaxy formation and evolution simulations in a cosmological context built on the MUFASA project (Davé, Thompson & Hopkins 2016) with updated physics for the modelling of black hole growth and feedback. SIMBA uses the meshless finite mass version of the GIZMO code (Hopkins 2015), assuming the Planck Collaboration XIII (2016) cosmological parameters of $\Omega_m = 0.3$, $\Omega_\Lambda = 0.7$, $\Omega_b = 0.048$, $H_0 = 68\,{\rm km\,s^{-1}\,Mpc^{-1}}$, $\sigma_8 = 0.82$, $n_s = 0.97$, and the star formation model of Schmidt (1959). This simulation explores the evolution of galaxies, black holes, and galactic gas over a wide redshift range. The molecular hydrogen content in SIMBA is calculated on the fly using the prescription provided in Krumholz & Gnedin (2011) based on the local gas metallicity and column density as follows:

$$f_{\rm H_2} = 1 - 0.75 \frac{s}{1 + 0.25s}. \qquad (1)$$

Where $f_{\rm H_2}$ is the molecular hydrogen mass fraction (mass of molecular hydrogen over the total gas mass of each gas cell) for each gas cell and $s$ is defined as

$$s = \frac{\ln(1 + 0.6\chi + 0.01\chi^2)}{0.0396 Z (\Sigma/{\rm M_\odot\,pc^{-2}})}, \qquad (2)$$

where $Z$ is the metallicity in solar units. The gas column density ($\Sigma$) needs to be estimated using the Sobolev (Sobolev 1960) approximation as $\Sigma = \rho h$ where $\rho$ is the mass density of gas and $h$ is the local density scale height in each gas cell in the simulation box, to be computed using $h = \rho/|\nabla\rho|$. In this equation, $\chi$ is a function of metallicity, defined as

$$\chi \approx 3.1 \left( \frac{1 + 3.1 Z^{0.365}}{4.1} \right). \qquad (3)$$

Having the molecular hydrogen mass fraction from the above equations, one can calculate the number density of molecular hydrogen $n_{\rm H_2}$ in each gas cell.

Stars only form in gas cells with $n_{\rm H} > 0.13\,{\rm cm^{-3}}$. The SFR in each gas cell of a galaxy is calculated using a stochastic star formation model from Schmidt (1959). In this model, the star formation rate is SFR $= \epsilon_* f_{\rm H_2} \rho / t_{\rm dyn}$, where $f_{\rm H_2}$, $\rho$, and $t_{\rm dyn}$ indicate molecular hydrogen mass fraction, gas mass density, and dynamical time-scale, respectively. The star formation efficiency per free-fall time is assumed to be $\epsilon_* = 0.02$ (Kennicutt 1998). Galactic star formation rates are computed as instantaneous SFRs from the gas elements. These values are also consistent with the SFRs calculated using young stellar particles, averaged over tens of Myr. SIMBA finds galaxies by applying a 6-dimensional Friends-of-Friends (6D-FoF) galaxy finder algorithm on the positions and velocities of all stars and gas cells with $n_{\rm H} > 0.13\,{\rm cm^{-3}}$. This FoF algorithm groups stars and gas particles into galaxies with a spatial linking length of 0.0056 times the mean interparticle distance and a velocity linking length equal to the local velocity dispersion. Haloes are found using a 3D-FoF method applied on dark matter particles with a linking length set to 0.2 times the mean interparticle spacing.

SIMBA models the star formation-driven galactic winds as decoupled two-phase winds with 30 per cent of the particles consisting of hot particles. The scaling relation of the mass loading factor with stellar mass comes from Anglés-Alcázar et al. (2017b) using the FIRE simulation (Hopkins et al. 2014). One of the main improvements of SIMBA compared to MUFASA is that black holes are seeded and grown on-the-fly and the feedback from black hole accretion contributes to galaxy quenching. SIMBA has two black hole accretion modes; torque-limited accretion for cold gas within the black hole accretion kernel ($T < 10^5$ K) using the model of Anglés-Alcázar et al. (2017a) based on Hopkins & Quataert (2011), and Bondi accretion (Bondi 1952) for hot gas ($T > 10^5$ K). The active galactic nuclei (AGN) feedback in SIMBA includes X-ray energy feedback based on the model of Choi et al. (2012) and a kinetic subgrid model consisting of radiative and jet modes. The radiative mode is used for high Eddington ratios ($f_{\rm Edd} = \dot{M}_{\rm BH}/\dot{M}_{\rm Edd} > 0.2$) when multiphase winds of molecular and warm ionized gas flow from AGNs with a speed of roughly $10^3$ km s$^{-1}$. The simulation begins to activate the jet mode feedback at lower Eddington ratios ($f_{\rm Edd} < 0.2$) and only becomes dominant at $f_{\rm Edd} < 0.02$. In this feedback mode, AGNs produce collimated jets of hot gas with a speed of roughly $10^4$ km s$^{-1}$. Additionally, these jets only exist in early-type galaxies with black hole mass of $M_{\rm BH} > M_{\rm BH,lim} = 10^{7.5} M_\odot$ to be consistent with observations. The velocities and temperatures of the bipolar kinetic feedback model are mostly taken from observation to reproduce the observed energy release on larger scales of tens of kpc (Davé et al. 2019). The jet mode AGN feedback has been shown to be the main mechanism responsible for galaxy quenching in SIMBA. The X-ray mode feedback is the complementary process to fully quench galaxies. However, the radiative mode does not contribute to galaxy quenching (Davé et al. 2019).

SIMBA tunes the feedback models to the galaxy stellar mass function (GSMF) at $z = 0$ since the GSMF is well measured over a reasonable redshift range (Davé et al. 2019). Assuming the observed GSMF from Bernardi et al. (2017), the combined Cosmic Assembly Near-infrared Deep Extragalactic Legacy Survey (CANDELS) and the FourStar Galaxy Evolution Survey (zFOURGE) data from Tomczak et al. (2014), and CANDELS data from Song et al. (2016), SIMBA has a GSMF in good agreement with these observations with the exception of a subtle overproduction of massive galaxies at $z < 2$. This inconsistency of the GSMF of massive galaxies between SIMBA and observations could be due to a number of observational biases or numeric uncertainties in SIMBA (see Davé et al. 2019).

In this work, we use 22 snapshots of the SIMBA full run simulation (m100n1024) over the redshift range of $0 < z < 4$. This simulation has been run in a box of $100 \, h^{-1}$ Mpc in length with $1024^3$ gas particles and $1024^3$ dark matter particles. The minimum gravitational softening length is $0.5 \, h^{-1}$ kpc. In this run, the gas and dark matter particle masses are $1.82 \times 10^7$ and $9.6 \times 10^7 \, M_\odot$, respectively. We apply a fixed lower stellar mass limit of $M_* > 10^9 \, M_\odot$ on our galaxy sample at all analyzed redshifts, which is the same as the lower mass limit of the sample that Tacconi et al. (2018) uses to find scaling relations between galaxy molecular gas masses, stellar masses, and star formation rates. Assuming this limit, we can compare our results with this scaling relation. Our SIMBA galaxy sample includes both centrals and satellites more massive than our mass limit at all redshifts. This data set includes 332 398 galaxies in total at all redshifts; and 5604 and 28 601 galaxies at $z = 4$ to $z = 0$, respectively. Galaxies at each redshift have evolved from progenitor galaxies at higher redshifts (that might not be within our stellar mass threshold). However, their stellar mass, environment, and other properties have also evolved over time and these galaxies can be considered as different data points in our sample. Considering the constant evolution of galaxies and their properties from $z = 4$ to $z = 0$, it is insightful to combine all galaxies at all redshifts into a single data set and consider redshift as an effective variable on the environmental trends as done in observational studies.

## 3 METHODS

In this section, we explain the methods used to investigate the effect of the environment on the star formation activity of galaxies. To achieve our goal, we study a number of important parameters in the star formation process of galaxies:

(i) Specific star formation rate [sSFR = SFR/$M_*$, unit (yr$^{-1}$)]: the rate of forming new stars per stellar mass.

(ii) Molecular hydrogen fraction ($f^*_{H_2} = M_{H_2}/M_*$): the mass fraction of the molecular hydrogen compared to stellar mass.

(iii) Star formation efficiency [unit (yr$^{-1}$)]: the efficiency of producing new stars in a galaxy, also the inverse of the time a galaxy takes to consume all of its molecular gas to form new stars ($t_{\text{depl}}$):

$$\text{SFE} = \text{sSFR}/f^*_{H_2} = \text{SFR}/M_{H_2} = 1/t_{\text{depl}} \quad (4)$$

Having these quantities from the simulation, we need an environment indicator to explore the relationship between the star formation process in galaxies and their large-scale environment.

### 3.1 Stellar mass density as an environment indicator

Among the various existing methods of measuring the environment of galaxies, we decided to investigate the relation between the stellar mass environmental density of the neighbour galaxies and the star formation activity of each galaxy. An advantage of using the stellar mass environmental density over using the number density of galaxies as an environment measure is that the stellar mass density in the vicinity of a galaxy can be directly linked to the luminosity density of neighbour galaxies on the same scale which is measurable from observations. Moreover, stellar mass density captures the underlying matter distribution better than number density which is susceptible to stochastic effects (Mo & White 1996; Wang et al. 2018). Furthermore, in both observations and simulations, the uncertainties in mass density are smaller than that in number density since small galaxies are numerous but small in mass (Tonnesen & Cen 2014). We also performed our analysis using the number density of neighbour galaxies that resulted in the same trends, but weaker. In this study, we calculate the stellar mass density by considering the stellar mass of each galaxy and all of its neighbour galaxies in a $1 \, h^{-1}$ Mpc grid. We choose $1 \, h^{-1}$ Mpc for our grid size because it represents the typical size of virialized galaxy clusters, so it is large enough to capture the superhalo environmental trends (Kauffmann et al. 2004; Muldrew et al. 2012). The general trends reported in this work remain the same when changing this parameter by a few megaparsecs. Our mass interpolation method is the Cloud In Cell (CIC) method that gives a fraction of the weight of each particle to all vertices of the cell it occupies based on the distance of the particle to each vertex (Birdsall & Fuss 1969). This method reduces density measurement errors compared to other mass assignment schemes. The dimensionless quantity environmental density (also known as overdensity or density contrast) is defined with the following equation:

$$\delta^*_{\text{gal}}(x) = (\rho(x) - \bar{\rho})\bar{\rho}. \quad (5)$$

Where $\rho(x)$ is the stellar mass density of galaxies in 3-dimensional space at point $x$ calculated on a $1 \, h^{-1}$ Mpc grid and $\bar{\rho}$ is the mean stellar mass density of the simulation box. PYLIANS3[1] (Villaescusa-Navarro et al. 2018) is our tool for this calculation which is a set of PYTHON libraries, written in PYTHON, CYTHON, and C to facilitate the analysis of numerical simulations. We use the defined environmental density ($\delta^*_{\text{gal}}$) as an indicator of the environment in this work. The results of this work remain the same if we consider dark matter mass rather than stellar mass density.

### 3.2 Error and scatter estimation

We use the median as a statistical tool to investigate the overall evolution of sSFR, $f^*_{H_2}$, and SFE with galaxy properties including stellar mass, redshift, and environmental density. The median is chosen as the statistical indicator of our data sample because it is robust against outliers and false measurements. In order to capture the entire behaviour of the data, we need to look at the measurement error of our statistical indicator and scatter of the data as well.

We use bootstrapping to measure the error of the median of sSFR, $f^*_{H_2}$, and SFE for our analysis. Bootstrapping is a resampling method useful for estimating bias and variance. In order to calculate a function in a data set with $N$ data points using bootstrapping, the first step is to generate $B$ resampled data sets ($B = 10\,000$ in our case) from the original data set each of size $N$ with replacement.[2] Then, one can calculate the function on each new data set and keep track of the statistical distribution of these values.

Furthermore, we need to quantify how the data points are scattered around the median values to avoid extracting trends only from median values. We use median absolute deviation (MAD) for measuring the scatter of data points around the median values. The MAD is defined

---

[1] https://pylians3.readthedocs.io/en/master/
[2] As a result of resampling with replacement in each iteration, some data points might appear more than once or not at all in the resampled data set.

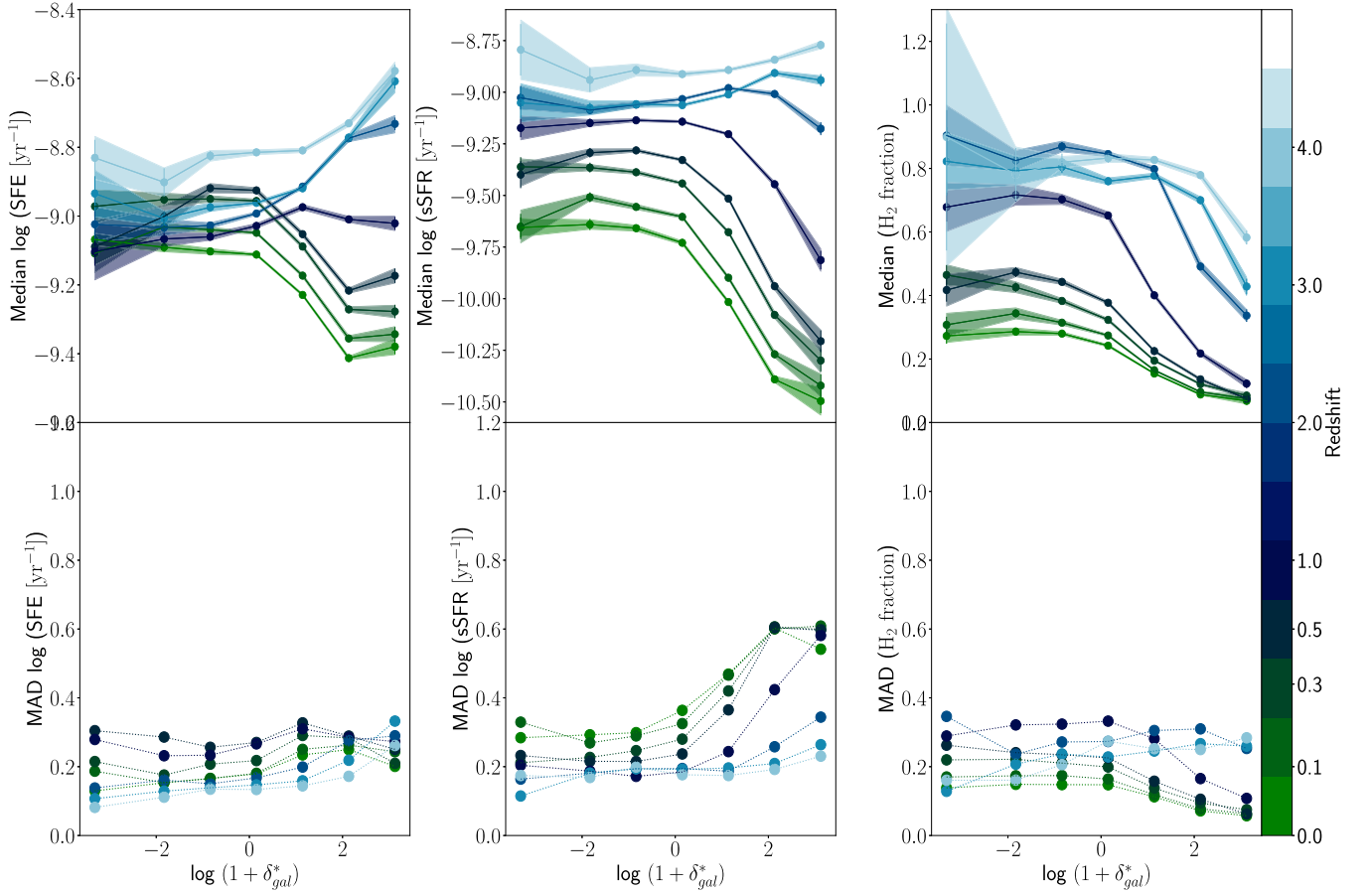

**Figure 1.** Top left: median galactic SFE(= sSFR/$f^*_{H_2}$) as a function of the environmental density, $\delta^*_{gal}$, at different redshifts, shown by the colour bar. Green colours show the present day at $z = 0$. In each environmental density bin, the median values and their uncertainties (shaded areas) are obtained from bootstrapping. Top middle: similar to the left panel for sSFR (=SFR/$M_*$) versus environmental density. Top right: same as the top left panel for the molecular hydrogen mass fraction $f^*_{H_2}(= M_{H_2}/M_*)$ versus environmental density. Bottom row: Median Absolute Deviation (MAD) of galaxy SFE, sSFR, and $f^*_{H_2}$ versus environmental density at different redshifts. The slope of the SFE–density curve is changing with redshift, showing the reversal of the SFE–density relation.

as the median of the absolute deviations from the median of the data for a data set $X_1, X_2,..., X_n$:

$$\text{MAD} = \text{median}(|X_i - \text{median}(X_i)|). \quad (6)$$

In our case, data points $X_i$ can be the log(SFE), log(sSFR), and $f^*_{H_2}$ of each galaxy with index $i$.

## 4 RESULTS

In the previous sections, we defined our galaxy sample, our methods, and the important parameters to investigate including specific star formation rate, molecular hydrogen mass fraction ($f^*_{H_2}$), star formation efficiency of galaxies, and environmental density ($\delta^*_{gal}$). In the remainder of the paper, we present and interpret our results based on galaxy evolution theories and other works on this topic, focusing on our central question: 'What is the effect of the Mpc-scale environment on the star formation activity of galaxies?'

### 4.1 The redshift evolution of star formation efficiency across galactic environments

In this subsection, we present the trends we detect between sSFR, $f^*_{H_2}$, SFE, and environmental density of galaxies in the SIMBA simulation. Fig. 1 presents the median star formation efficiency, median specific star formation rate, and the median molecular hydrogen mass fraction of galaxies as functions of the environmental density (as defined in Section 3.1) in the redshift range of $0 < z < 4$ in the top row. We only show eight redshift snapshots in this figure for visual clarity. The star formation efficiency is shown in the top left panel of Fig. 1, indicating that at high redshifts ($z > 1$) galaxies in denser regions form stars more efficiently than galaxies in low-density regions, for instance by ∼0.2 dex at $z = 4$. However, at lower redshifts ($z < 1$) galaxies in denser regions form stars less efficiently than galaxies in low-density regions, for instance by ∼0.25 dex at $z = 0$.

Investigating the environmental density dependence of the specific star formation rate and the molecular hydrogen mass fraction is insightful to interpret the visible trends for SFE. The top middle panel of Fig. 1 presents the median sSFR values for galaxies as a function of the environmental density for different redshift snapshots. At high redshifts ($z > 1$), the sSFR–density curves are almost flat on average, with subtle variations in the densest bins. The slope of the curves becomes negative towards lower redshift ($z < 1$), meaning that at these redshifts galaxies in denser regions have lower sSFR values compared to galaxies in low-density regions, for example by ∼0.9 dex at $z = 0$. Finally, the top right panel of Fig. 1 shows the environmental density dependence of the molecular hydrogen mass fraction of galaxies at different redshift snapshots. This plot shows that the molecular hydrogen mass fraction of galaxies decreases with environmental density at all redshifts by ∼0.2 for $z = 4$ and ∼0.15 for $z = 0$.

Considering all of the trends shown in Fig. 1, one can see that at high redshift ($z > 1$), the increase of SFE with environmental density arises from the decrease of $f_{H_2}^*$ with environmental density. However, at low redshift ($z < 1$), the decrease of sSFR with environmental density seems to be stronger than the decrease of $f_{H_2}^*$ with environmental density, leading the SFE = sSFR/$f_{H_2}^*$ to decrease with environmental density as well. Overall, the slope of the SFE $- \delta_{gal}^*$ curves is negative at low redshifts ($z < 1$) and positive at high redshifts ($z > 1$). We can call this change in the slope of the SFE $- \delta_{gal}^*$ at $z \sim 1$ a 'reversal' of SFE–density relation [analogous to the reversal of SFR–density relation introduced in Elbaz et al. (2007) and Popesso et al. (2015)].

The bottom row of Fig. 1 presents the MAD of SFE, sSFR, and the molecular hydrogen mass fraction versus the environmental density of galaxies with different colours indicating the different explored redshifts. MAD is a measure of the scatter in the data. The scatter of galaxy sSFR, plotted in the bottom middle panel of Fig. 1, increases with environmental density at all redshift snapshots. We think it results from high galactic activity in dense regions, including merger and AGN feedback, compared to lower density regions. However, the scatters of SFE and $f_{H_2}^*$ do not directly depend on the environmental density.

In order to investigate the reversal of the SFE–density relation, we look at the redshift evolution of the SFE–density relation directly in Fig. 2. These plots show SFE, sSFR, and $f_{H_2}^*$ as functions of redshift for different overdensities, shown by different colours. The redshift evolution of SFE is plotted in the left panel of Fig. 2, suggesting a reversal of SFE–density relation around $z \sim 1$. At high redshift ($z = 4$), galaxies in dense regions form stars more efficiently than galaxies in low-density regions by ∼0.3 dex. While in the local universe ($z = 0$), galaxies in dense regions form stars less efficiently than those in low-density regions by ∼0.3 dex.

The top right panel of Fig. 2 shows that in the high-redshift universe in SIMBA ($z \sim 4$), galaxies in denser regions (darker curves) have slightly higher median sSFR (∼0.2 dex) than galaxies in low-density regions (lighter curves). However, this trend changes at lower redshifts in a way that at $z \sim 0$, galaxies in high-density regions have much lower median sSFR (∼0.9 dex) than galaxies in low-density regions. Based on this panel, one can argue that we have a reversal of the sSFR–density relation with a turning point around $z \sim 2.5$. The drop in the sSFR-redshift curves occurs earlier in more dense regions. Hence, this reversal in the sSFR–density relation reflects that galaxy quenching happens earlier in high-density regions compared to low-density regions. The bottom right panel of Fig. 2 shows the redshift evolution of the molecular hydrogen mass fraction for different environment densities. From this plot, it is clear that galaxies in the two densest bins on average have larger molecular hydrogen mass fractions at high redshift compared to the low-redshift snapshots. The same trend can be seen for galaxies in low-density regions after a certain redshift. Furthermore, galaxies in denser regions have a lower molecular hydrogen mass fraction compared to those in low-density regions at all redshifts. Consequently, the reversal of SFE–density relation mainly arises from the redshift evolution of sSFR.

**4.2 Redshift evolution of SFE–density relation: environment or mass dependence?**

We detect an environmental dependence for the redshift evolution of the star formation activity of galaxies in Fig. 2. However,

environmental density may just be one of the many factors affecting the star formation activity of galaxies. According to observed scaling relations, other factors like the stellar mass of galaxies may also affect their star formation efficiency (Peng et al. 2010). Moreover, the correlation between massive galaxies and dense environments may bias the interpretation of environmental trends (Bolzonella et al. 2010; Darvish et al. 2015; Bahé et al. 2017). Furthermore, if an environmental dependence persists after controlling for stellar mass, we can conclude that the environment is indeed one factor leading to the trends reported in the last section. In this section, we divide our galaxy sample into three stellar mass bins in order to distinguish between the effect of mass and the effect of the environment on the star formation activity of galaxies.

Fig. 3 shows the dependence of the SFE, sSFR, and the molecular hydrogen mass fraction of galaxies on the environmental density for different redshifts and galaxy stellar masses. Each column in this figure demonstrates galaxy properties in a specific stellar mass bin, starting from galaxies with $9 < \log \frac{M_*}{M_\odot} < 9.5$ in the left column (containing $N_{gal} = 147\,133$ galaxies), to galaxies with $9.5 < \log \frac{M_*}{M_\odot} < 10$ in the middle column (with $N_{gal} = 98\,083$), and galaxies with $10 < \log \frac{M_*}{M_\odot}$ in the right column (with $N_{gal} = 87\,182$). We have three stellar mass bins, 22 redshift bins, and five environmental density bins, generating 330 data points in each row of Fig. 3. Information on the number of galaxies in each bin is provided in Table A1.

The top row panels of Fig. 3 explore the redshift evolution of SFE. In the top left and top middle panels (low- and intermediate-mass bins) we can see that for the galaxies in the local universe ($z \sim 0$), those in denser regions have ∼0.3 dex lower SFE values than those in low-density regions. However, we do not see a significant dependence on the environmental density for low- and intermediate-mass galaxies in the high-redshift universe. Furthermore, the effect of the environmental density on the SFE of massive galaxies (top right panel) is much less prominent than in lower mass galaxies. In fact, only galaxies in the densest bin have ∼0.2 dex higher SFE than other galaxies at $z = 0$. Comparing the top row of Fig. 3 with the left panel of Fig. 2, we can see that at $z \sim 4$, the visible trend in Fig. 2 is not an effect of the environmental density, but it mainly arises from the large range of SFE values in different mass bins. The overall trend in the low-redshift universe ($z < 1$) of Fig. 2 is seen in the low- and intermediate-mass bins of Fig. 3, but not in its most massive bin. Hence, the environmental density has a small effect on the star formation efficiency of low- and intermediate-mass galaxies in the low-redshift universe.

The middle row of Fig. 3 presents the redshift evolution of sSFR. Similar to the SFE plots in the top row, an environmental density dependence of sSFR in the local universe ($z \sim 0$) is noticeable (∼0.5 dex) in the low- and intermediate-stellar mass galaxies. However, these low- and intermediate-mass galaxies do not show a density dependence of their sSFR in the high-redshift universe ($z \sim 4$). For the most massive galaxies, an environmental dependence is only discernible between redshifts $z \sim 1$ and $z \sim 3$. At this redshift range, galaxies in denser regions have lower sSFR values than those in low-density regions. There is no obvious trend seen for these galaxies in the local universe ($z \sim 0$) or the highest redshift bins ($z > 3$). By comparing the middle row of Fig. 3 with the top right panel of Fig. 2, one can argue that the overall trend at the local universe ($z \sim 0$) is primarily driven by lower mass galaxies ($9 < \log \frac{M_*}{M_\odot} < 10$).

The redshift evolution of molecular hydrogen mass fraction is illustrated in the bottom row of Fig. 3 for galaxies at different stellar mass bins, and environment overdensities. The most significant trend in these plots is that all galaxies in the densest bins have the lowest hydrogen mass fraction at almost all redshifts explored in this work ($0 < z < 4$). Moreover, more massive galaxies have lower hydrogen fractions, on average, which is consistent with the trend seen in Davé et al. (2020). The molecular hydrogen mass fraction shows the most prominent dependence on environmental density, more than that of the sSFR and the SFE (Figs 2 and 3).

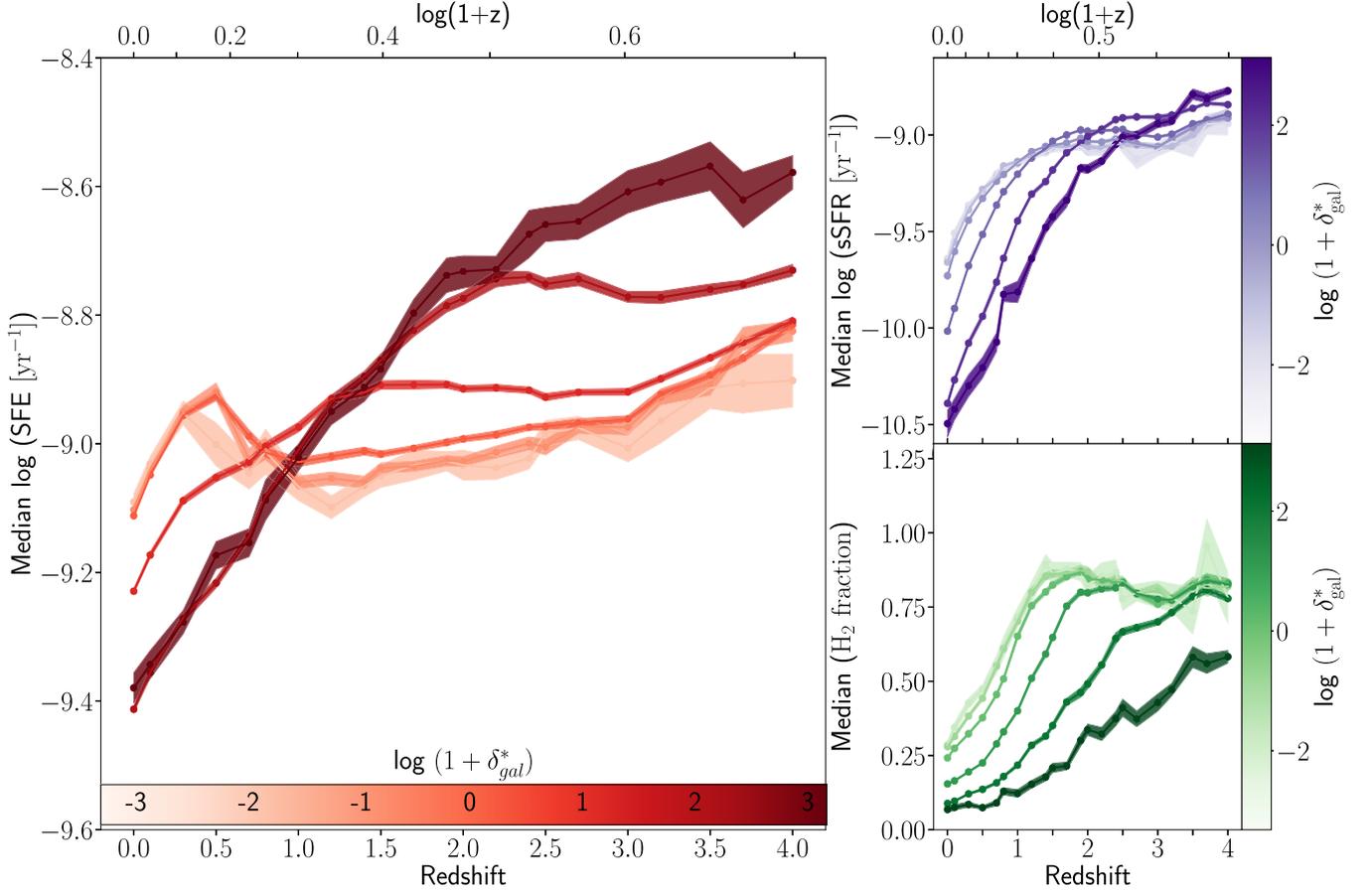

**Figure 2.** SFE, sSFR, and molecular hydrogen mass fraction as functions of redshift for different environmental densities, shown by colours. The reversal point of the SFE–density and sSFR–density relations are at $z \sim 1$ and $z \sim 2.5$, respectively.

In conclusion, our analysis shows a slight dependence of the star formation efficiency of galaxies in SIMBA at low- to intermediate-mass bins ($9 < \log \frac{M_*}{M_\odot} < 10$) at low-redshifts ($z < 1.5$). However, we do not detect a reversal of SFE–density relation in Fig. 3 in each mass bin, meaning that this is a pure effect of the stellar mass of galaxies and not their environment.

### 4.3 Scaling relations for gas depletion time

We fit scaling relations to the galaxy parameters to examine the statistical significance of the trends observed in Figs 2 and 3. Tacconi et al. (2018) fits the depletion time ($t_{\rm depl}$) as a function of redshift ($z$), specific star formation rate (sSFR), stellar mass ($M_*$), and galaxy's half-light radius ($R_{\rm h}$) in the rest-frame optical band 5000 Å with the following equation:

$$\log t_{\rm depl}({\rm Gyr}) = A_t + B_t \log(1+z) + C_t \log(\delta MS) + D_t (\log M_* - 10.7) + E_t \log(\delta R_{\rm h}). \quad (7)$$

Where $\delta MS = {\rm sSFR}/{\rm sSFR}_{({\rm MS},z,M_*)}$ in which ${\rm sSFR}_{({\rm MS},z,M_*)}$ is the specific star formation rate of the star-forming main-sequence galaxies, defined in Speagle et al. (2014) with the following equations:

$$\log \frac{{\rm sSFR}_{({\rm MS},z,M_*)}}{{\rm Gyr}^{-1}}) = 9 - (6.51 - 0.11 \times \frac{t_{\rm c}}{\rm Gyr}) + (-0.16 - 0.026 \times \frac{t_{\rm c}}{\rm Gyr}) \times (\log \frac{M_*}{M_\odot} + 0.025), \quad (8)$$

where

$$\log \frac{t_{\rm c}}{\rm Gyr} = 1.143 - 1.026 \times \log(1+z) - 0.599 \times \log^2(1+z) + 0.528 \times \log^3(1+z), \quad (9)$$

Furthermore, $\delta R_{\rm h} = R_{\rm h}/R_{\rm e0}(z, M_*)$, where $R_{\rm e0}(z, M_*)$ is the average half-light radius of the star-forming population, defined in van der Wel et al. (2014) as $R_{\rm e0} = 8.9\,{\rm kpc}(1+z)^{-0.75}(M_*/5 \times 10^{10}{\rm M_\odot})^{0.23}$.

Tacconi et al. (2018) fits equation (7) to the Plateau de Bure High-z Blue Sequence Survey (PHIBBS) survey, consisting of a large sample of 1444 star-forming galaxies in the redshift range of $0 < z < 4$, a stellar mass range of $9 < \log \frac{M_*}{M_\odot} < 11.8$, and the star formation rate relative to the main sequence in range of $10^{-1.3} < \delta MS < 10^{2.2}$. We show the best-fitting function in Fig. 4 and its parameters are:
$A_t = +0.09$, $B_t = -0.62$, $C_t = -0.44$, $D_t = +0.09$, $E_t = +0.11$

Krumholz & Dekel (2012) and Hunt et al. (2016) argue that metallicity affects the star formation activity of galaxies as well. Following this idea, we include this parameter in our study on SIMBA. In this work, we modified equation (7) to include two additional terms to fit for the environmental density ($\delta^*_{\rm gal}$) and the galaxy mass-

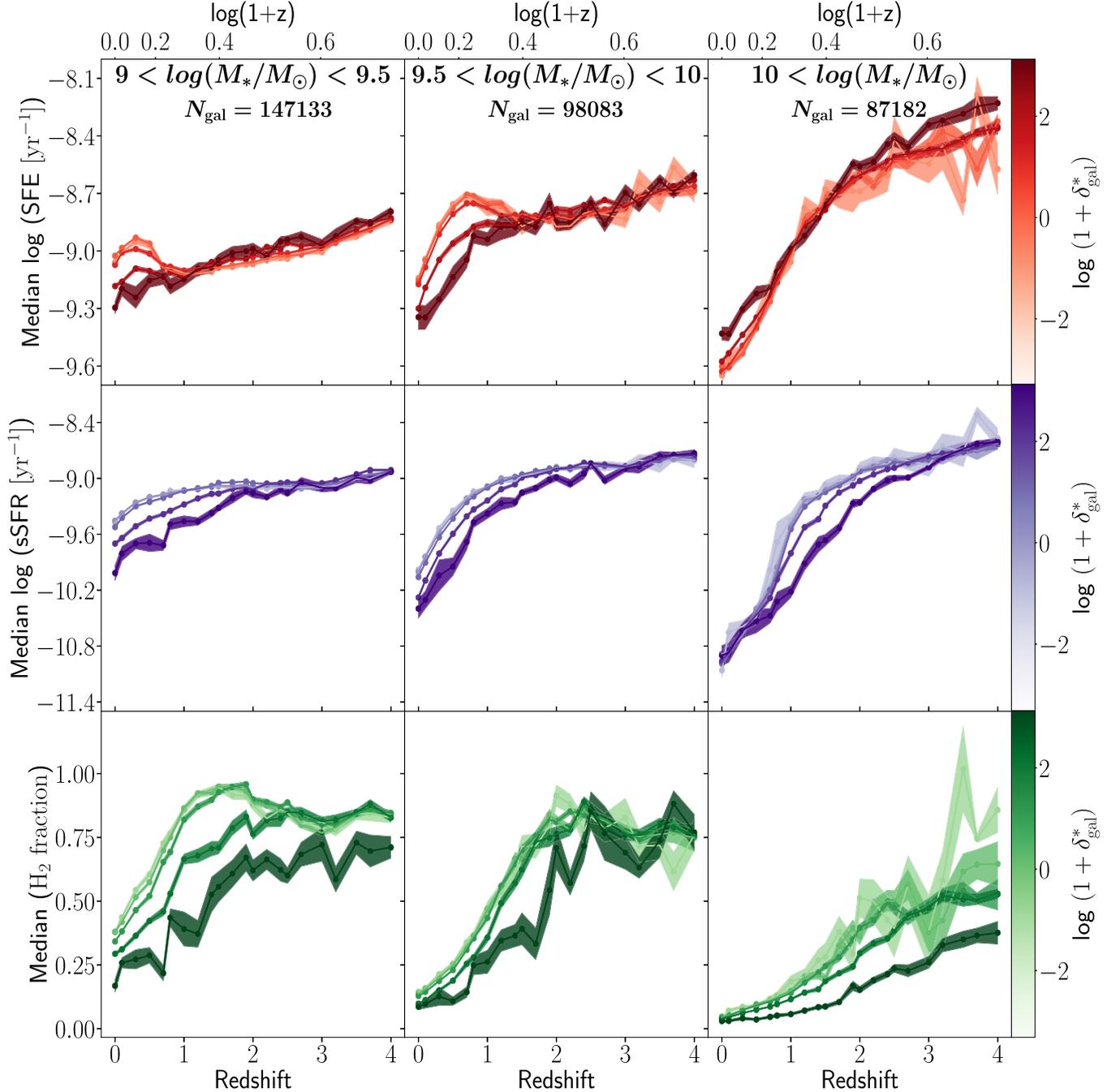

**Figure 3.** *Top:* median SFE as a function of redshift for different stellar masses and environment overdensities, calculated using bootstrapping as described in Section 3.2. Each column shows a stellar mass bin and colours represent the environmental density bin. Shaded regions show the standard deviation of bootstrapped median values. $N_{\rm gal}$ in each column presents the number of all galaxies at all redshifts used to derive the medians. *Middle:* same as the top row, but for the specific star formation rate (sSFR). *Bottom:* same as the top row for molecular hydrogen mass fraction $f^*_{H_2}$. The environmental dependence of SFE is mostly prominent for low- and intermediate-mass galaxies at low redshift.

weighted metallicity (Z):

$$\log t_{\rm depl}({\rm Gyr}) = A + B \log(1+z) + C \log {\rm sSFR} + D \log M_* + E \log(R_{\rm hm}) + F \log(Z) + G \log(1+\delta^*_{\rm gal}). \quad (10)$$

We use the half-mass radii ($R_{\rm hm}$) of galaxies from the SIMBA simulation in this equation. We evaluate the significance of each term by performing linear regression on equation (10). For this fitting function, the 'coefficient of determination' $R^2$ is 0.866 2025. This parameter ranges from 0 to 1 with higher values showing statistically better fits.[3] The coefficients, their standard errors, and their 95 per cent confidence intervals for this fitting function are presented in Table 1. Excluding the environmental density leads to a slightly poorer fit with $R^2 = 0.866\,1609$. Excluding the metallicity parameter from the main fit also gives a weaker fit with $R^2 = 0.865\,2847$. This is consistent with the recent observational studies, showing the environmental dependences of the gas-phase metallicity of galaxies (Chartab et al. 2021; Calabrò et al. 2022).

---

[3]Coefficient of determination ($R^2$) is a statistical measure for the goodness of fit, calculated as $R^2 = 1 - \frac{\sum_i (X_i - X_{\rm pred,i})^2}{\sum_i (X_i - \overline{X})^2}$, where $X_i$ are the data set values, $X_{\rm pred,i}$ are the predicted values by the model, and $\overline{X}$ is the average value of the data set.

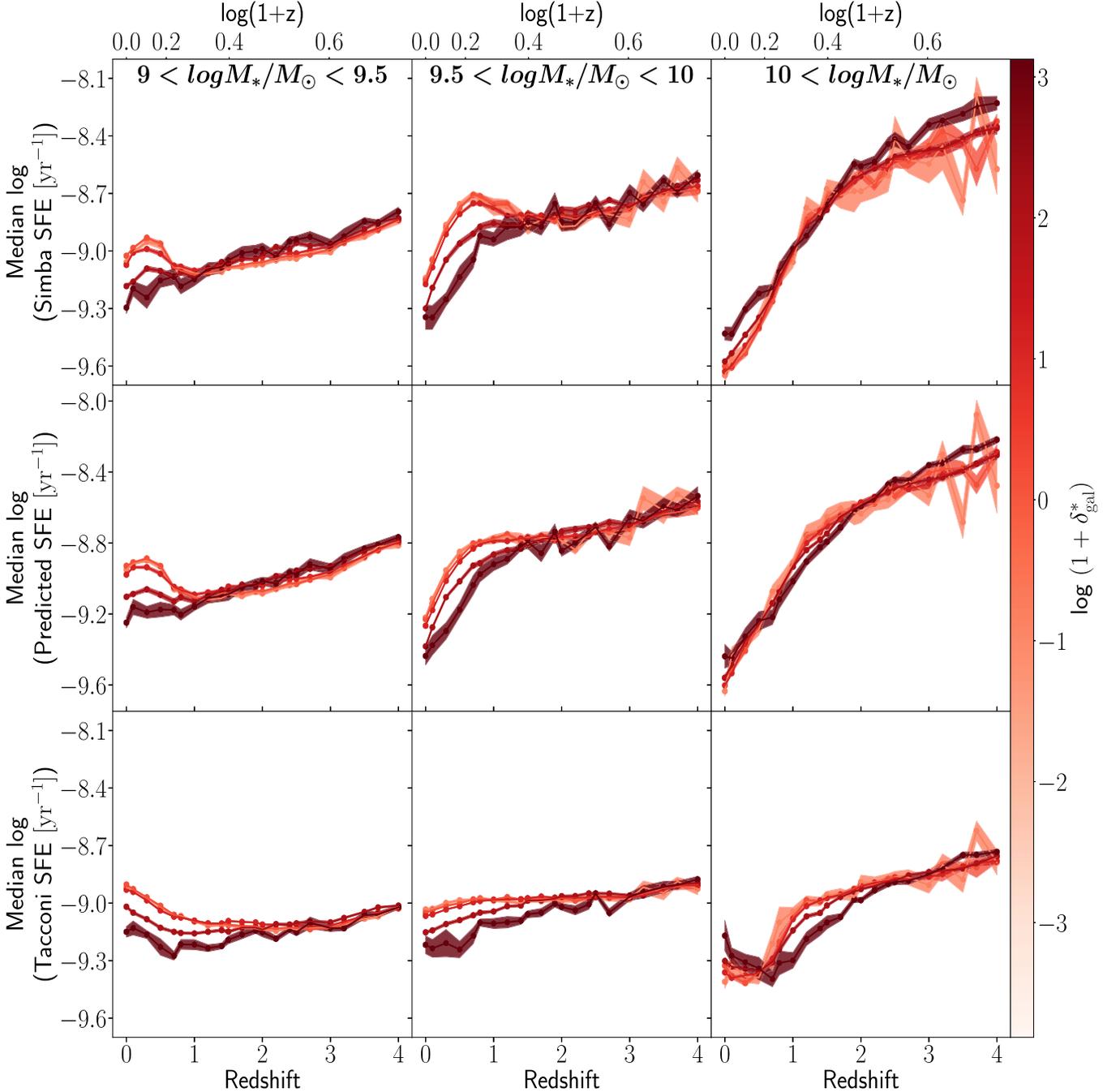

**Figure 4.** Cosmic evolution of SFE for galaxies with different stellar masses. *Top*: the actual SFE inferred from SIMBA. *Middle*: the SFE predicted by the best scaling relation from this work (Table 1). *Bottom*: the SFE computed from the observational scaling relation by Tacconi et al. (2018) following the procedures described in Section 4.3. Colours show environmental density. The environmental dependence of SFE (seen in Fig. 3) is reproduced using our scaling relation.

Fig. 4 presents a comparison between the actual star formation efficiency of SIMBA galaxies, the SFE predicted by our model, and the SFE predicted by the Tacconi et al. (2018) scaling relation.

As expected, the environmental dependence of galactic SFE seen in SIMBA at different redshifts, stellar mass, and environmental density bins is captured in the predicted SFEs from our scaling relation. Interestingly, the SFE values predicted by the Tacconi et al. (2018) relation show an environmental dependence even though environmental density is not an explicit parameter in their scaling relation. This might be due to the environmental dependence of other fit parameters, for instance, sSFR and size. This comparison illustrates that there are trends not fully captured in the Tacconi et al. (2018) parametrization without the $(1 + \delta^*_{\rm gal})$ term. Fig. 5 shows a relative comparison between the actual SIMBA SFE values and the SFE predicted by our scaling relation. The scaling relation presented in equation (10) and Table 1 yields SFE values within 30 per cent of the Simba values. The predicted values follow the general trends of our SFE-redshift curves in most of the bins.

## 5 DISCUSSION

Mass and environment are the two determinants suggested in the literature for the evolution of galaxies (Peng et al. 2010, 2012; Darvish et al. 2016). The effects of these factors on star formation activity have been studied from different viewpoints and using different methods. In this section, we discuss our results in the context of the literature findings.

### 5.1 Impact of the stellar mass on star formation

The stellar mass of a galaxy is an important indicator of its evolutionary path, including its past and ongoing star formation activity. For instance, more massive galaxies in SIMBA have more massive black holes consistent with observations (Davé et al. 2019). According to the black hole growth model of SIMBA, the accretion rate of black holes increases with their mass (Davé et al. 2019). Therefore, galaxies with larger stellar masses and corresponding higher black hole masses have higher black hole accretion rates. Moreover, Thomas et al. (2019) show that SFR increases with the black hole accretion rate for star-forming (SFR > 1 M$_\odot$ yr$^{-1}$) galaxies in SIMBA at redshift $0 < z < 5$. They suggest this arises from the common gas reservoir used for star formation and torque-limited black hole growth mode.

The AGN jet feedback in SIMBA is the main quenching responsible for galaxies with black hole mass $M_{BH} > 10^{7.5}$ M$_\odot$ which corresponds to stellar mass $M_* > 10^{10}$ M$_\odot$ (Thomas et al. 2019). Consequently, the most massive galaxies we investigate in this work are significantly affected by the AGN jet feedback which might be one of the reasons their SFE and sSFR drop more rapidly than lower-mass galaxies (Fig. 3). Since AGN jet particles in SIMBA are decoupled until they reach outside the galaxy (tens of kpc), they cannot directly entrain gas in the interstellar medium (ISM) of the galaxy (Davé et al. 2019). However, the AGN jets in SIMBA are implemented to heat up the gas in the circumgalactic medium (CGM) of galaxies which truncates the accretion of cold gas from the CGM into the galaxy (Appleby et al. 2021). We do not investigate the impacts of AGN in this work, however, a potential trend could be as follows: since more massive galaxies tend to reside in more dense regions (Bahé et al. 2017), their strong AGN jet feedback can disturb the cooling and accretion of the CGM gas, resulting in less fuel for star formation and a drop in the sSFR of these galaxies compared to those in lower-density regions (the right column of Fig. 3). The suppression of cooling as a result of AGN feedback (Dubois et al. 2010) can cause a lower fraction of dense gas and thus lower SFE in these galaxies compared to lower-mass galaxies. The faster drop of median SFE and median sSFR of massive galaxies compared to lower mass galaxies in Fig. 3, can partially cause the reversal of median SFE–density and median sSFR–density relations we show in Fig. 2.

It is commonly known that mass quenching is more effective in more massive galaxies at higher redshifts while environment quenching is mostly effective in lower mass galaxies at lower redshifts (Peng et al. 2010; Darvish et al. 2018a). Recent observational studies find quenched massive galaxies (with mass completeness limit of $\log M_*/M_\odot > 10$–11, comparable to our most massive bin) in high-redshift protoclusters ($1.4 < z < 3.3$) (Zavala et al. 2019; Alberts et al. 2022; McConachie et al. 2022; Ito et al. 2023; Mei et al. 2023). Since galaxy protoclusters are not fully virialized in this phase, the effect of environment quenching might not be as significant as the effect of mass quenching. Assuming that these recently reported protoclusters are overdense regions at the observed redshifts, they could be in line with the trends seen in Fig. 3, showing that the most massive galaxies of SIMBA in dense regions have lower sSFR compared to those in low-density regions at $0.5 < z < 3$. Measuring the quenched fraction of observed galaxies outside protoclusters would be useful to test the validity of the SIMBA results.

Zavala et al. (2019) attributed the observed quenched massive galaxies in protoclusters to the early phases of environment quenching. They argue that this process is due to the environment because of the high fraction of quenched galaxies in overdense regions. They use the Atacama Large Millimeter Array (ALMA) Band 6 observations of the dust continuum to investigate the SFE and gas fractions (defined as the ratio of the ISM gas mass to the ISM gas mass plus stellar mass) of 68 member galaxies of two massive protoclusters at $2 < z < 2.5$. Their sample spans a stellar mass range of $9 < \log \frac{M_*}{M_\odot} < 11.6$. While they find, on average, similar scaling relations between the most massive galaxies ($\log \frac{M_*}{M_\odot} > 10$) in protoclusters and those from the field, they argue that a discernible fraction of the lower mass galaxies might have enhanced star formation efficiencies compared to field galaxies. Their observed trend for massive galaxies is consistent with our detected trends presented in Fig. 3, showing no environmental dependence for SFE for massive galaxies. However, their low-mass galaxy sample ($\log \frac{M_*}{M_\odot} < 10$) is incomplete and cannot be used to extract any trends. Our SIMBA results predict lower SFE in dense regions for low-mass galaxies (Fig. 3). Current ALMA observations do not provide constraints for low-mass galaxies and deeper observations would yield useful constraints on simulations and the trends they predict.

### 5.2 Impact of the environment on star formation

Our analysis shows that at low redshift ($z < 1.5$), lower-mass galaxies with $9 < \log \frac{M_*}{M_\odot} < 10$ in the SIMBA simulation that reside in denser regions have, on average, lower SFEs, and lower sSFRs compared to galaxies in low-density regions (Fig. 3). This trend has been seen in other observational and simulation works as well and usually is attributed to environmental quenching caused by different environmental processes.

More mergers in dense regions could be a reason for environmental quenching. Rodríguez Montero et al. (2019) study mergers and the resulting starburst and quenching at $0 < z < 2.5$ in SIMBA. They find in merger events that the SFR can increase up to 2–3 times the SFR of normal star-forming galaxies. They show that this jump in SFR for low-mass galaxies ($\log \frac{M_*}{M_\odot} < 10.5$) is due to the higher molecular gas content in the final galaxy. While, for more massive galaxies ($\log \frac{M_*}{M_\odot} > 10.5$), this jump arises from higher SFE because of more dense molecular gas in the galaxy after the merger. Since merger events are more abundant in denser regions, this SFR jump for mergers could result in an enhancement of SFR of galaxies in denser regions, in contradiction to our trends shown in the middle row of Fig. 3. However, the SFR jump caused by mergers in Rodríguez Montero et al. (2019) makes up a very little part (a couple of per cents) of the overall cosmic SFR of SIMBA and its effect cannot be significantly detected using our measures. Consequently, the SFR enhancement in dense regions caused by mergers is much smaller than the SFR drop in denser regions we find in Fig. 3 and it is not discernible in our work.

**Table 1.** Depletion time of galaxies as functions of redshift, stellar mass, specific star formation rate, stellar mass overdensity, metallicity, and galaxy size. In this table, the 'Coefficient' column is the predicted coefficient for the variable specified in the first column (defined as A–G in equation 10). 'STD' and 'Confidence Interval' columns are the standard deviation and the 95 per cent confidence interval of the coefficient, respectively.

| Variable | Coefficient | STD | Confidence interval |
| --- | --- | --- | --- |
| Const. | −1.775 | 0.016 | [−1.806, −1.744] |
| $1+z$ | −0.429 | 0.002 | [−0.433, −0.426] |
| sSFR | −0.450 | 0.001 | [−0.451, −0.449] |
| $M_*$ | −0.320 | 0.001 | [−0.322, −0.318] |
| $R_{hm}$ | 0.776 | 0.002 | [0.772, 0.779] |
| $Z$ | −0.078 | 0.002 | [−0.081, −0.075] |
| $1+\delta^*_{gal}$ | 0.003 | 0.0003 | [0.002, 0.003] |

Darvish et al. (2016) finds that the environment and mass quenching depend on each other in a way that environment quenching happens more efficiently for massive galaxies than lower mass galaxies and mass quenching occur more efficiently in dense regions. They suggest mergers as the main cause of this finding. They investigate the relationship between the galactic environment, stellar mass, and star formation activity for 73 481 galaxies in the Cosmic Evolution Survey (COSMOS) in the redshift range of $0.1 < z < 3.1$. Their mass-complete galaxy sample has a $K_s$-band magnitude limit of $K_s < 24$ and a stellar mass range of $9.14 < \log \frac{M_*}{M_\odot} < 11.5$. They measure the SFRs and stellar masses using a spectral energy distribution (SED) template fitting to the available UV, optical, and mid-infrared data. The limiting stellar mass of this data set grows with redshift, such that in the redshift range of $0.1 < z < 0.5$ the data set is complete for galaxies with stellar mass $9.14 < \log \frac{M_*}{M_\odot}$, while this limit grows to 9.97 at $1.5 < z < 3.1$. Darvish et al. (2016) uses the Voronoi tessellation method to calculate the environmental density of galaxies. They show that at $z < 1$ the median SFR and median sSFR decrease with increasing density, while they become independent of redshift at $z > 1$. They argue that environmental quenching is the dominant quenching process at $z < 1$. At $z > 1$, mass quenching is the dominant process that likely arises from stellar feedback.

In general, a direct and precise comparison of observations and simulations is not feasible due to the different nature of these data. Furthermore, the methods used to measure the galactic parameters, including SFR, molecular hydrogen, and environment in our work are different from Darvish et al. (2016). Nonetheless, it is still instructive to perform a qualitative comparison. Keeping these points in mind, we look at SFR and sSFR as functions of density for different redshifts in our work and Darvish et al. (2016) in Fig. 6 for comparison. We apply the same mass completeness limits on our galaxies as their work (table 1 of Darvish et al. 2016) for a fair comparison. Fig. 6 shows that the density dynamic range of the Darvish et al. (2016) data is smaller compared to our data (by ∼4 dex), but the SFR and sSFR dynamic ranges of their galaxies are larger than our galaxies (by 1–2 dex each). Hence, Darvish et al. (2016) reports stronger trends than our work. The first row of Fig. 6, showing SFRs from both works, suggests that both data sets follow qualitatively the same trends. The median SFR of SIMBA increases towards $z > 3$ and the highest density bins ($\log(1+\delta^*_{gal}) > 1$), a parameter space beyond that probed by Darvish et al. (2016). Based on table 1 from Darvish et al. (2016), the mass completeness limit for their galaxies with redshifts higher than 1.1 is $10^{9.93}$ M$_\odot$, so their sample at high redshift ($z > 1$) just covers the most massive galaxies from our sample and is incomplete in detecting what we classify as lower mass galaxies in this work. Since we use the same mass limits as Darvish et al. (2016) in Fig. 6, the trends we see in this figure are weaker than the trends found in Fig. 1. Although there are noticeable differences in the dynamic range in median SFR between our work and Darvish et al. (2016), we find qualitatively similar trends for star formation as a function of the environment for all of our galaxies, on average.

A bump is noticeable in the SFE and molecular hydrogen mass fraction of our lower mass galaxies ($\log \frac{M_*}{M_\odot} < 10$) at $z < 1.5$ that reside in lower density regions in Fig. 3. A previously quenched galaxy can 'rejuvenate' if it experiences a merger event or accretes a lot of cold gas from an external gas reservoir. In this case, the gas content and star formation activity of the galaxy would increase after a drop caused by quenching. However, based on Lorenzon et al., in preparation, the rejuvenation event among the quenched and post-starburst SIMBA galaxies is less than 10 per cent at each redshift and is independent of environments. Therefore, the rejuvenation events cannot explain the detected trends in our work.

The 'starvation' of satellite galaxies happens when they fall into galaxy clusters with hot intracluster medium (ICM). As a result, they are prevented from accreting enough cold gas from their surrounding which leads them to quench after finishing their gas content. This theory has been tested by van de Voort et al. (2017)

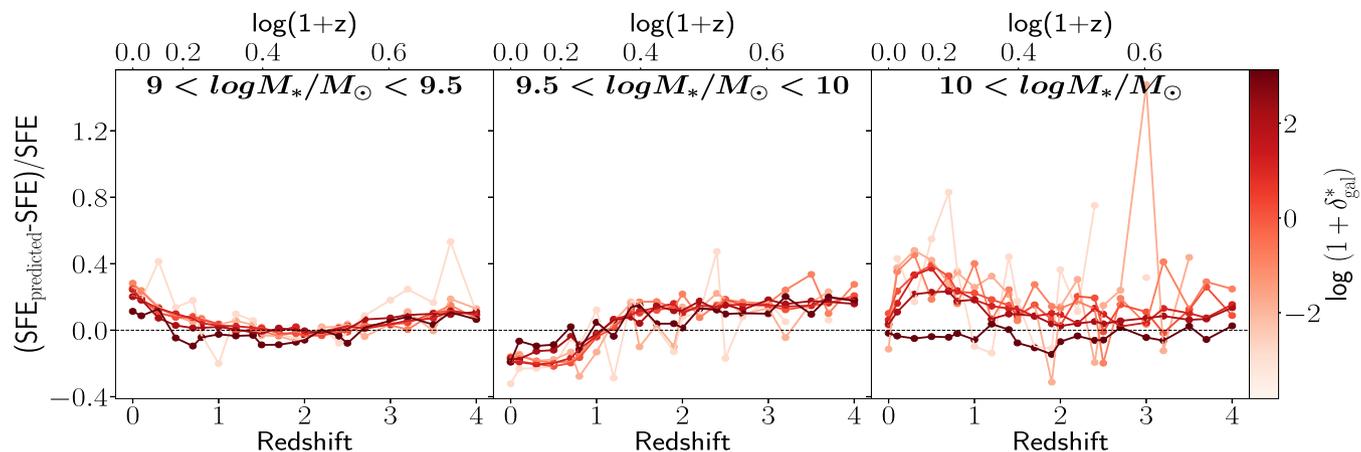

**Figure 5.** The relative difference of the predicted SFE using our scaling relation (SFE as a function of $z$, sSFR, $M_*$, $R_{hm}$, $Z$, and $\delta^*_{gal}$) and the actual SFE values as a function of redshift for different stellar mass bins and environmental density bins. The model predictions are close to the data.

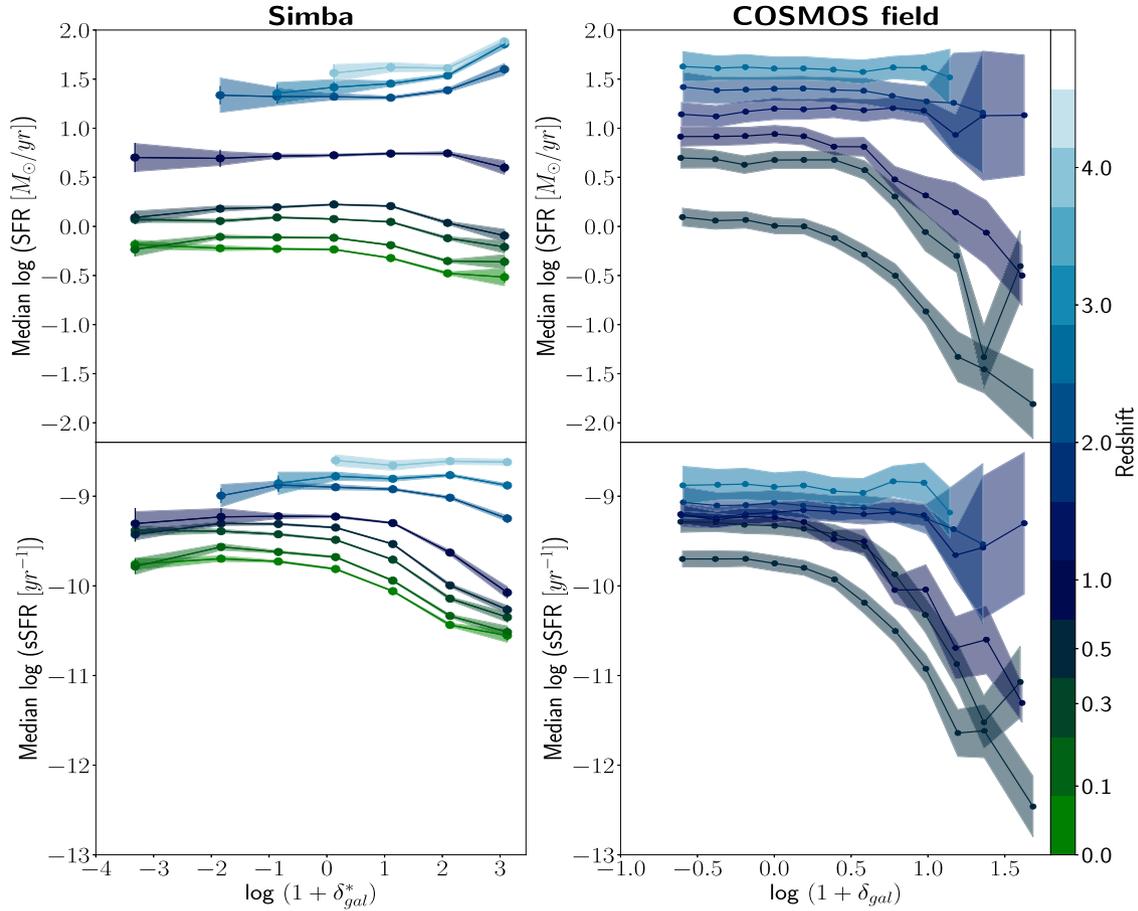

**Figure 6.** Comparison of the environmental dependence of star formation activity in SIMBA and the observational study based on the COSMOS survey presented in Darvish et al. (2016). Both data sets have the same stellar mass lower limits. *Left:* SFR (top) and sSFR (bottom) as functions of the environmental density at different redshifts, shown by the colour bar for SIMBA. *Right:* same as the left column for Darvish et al. (2016) data (the right column is a reproduction of figure 1 of Darvish et al. 2016). For visual purposes, the *x*-axes of the plots in the right column are more zoomed-in than the left column. SIMBA is qualitatively reproducing the observed trends in the COSMOS field, albeit SIMBA shows a smaller dynamic range in SFR and sSFR overall.

using the EAGLE cosmological simulation. They study galaxies with $\log \frac{M_*}{M_\odot} > 8$ at redshifts $0 < z < 2$ with the environment defined as the local 3-dimentional number density of galaxies up to the 10th nearest neighbour. They find a strong suppression of gas accretion rate in dense environments, most effective on satellites at low redshifts. Starvation may thus be a plausible explanation for the low star formation activity of galaxies in dense regions compared to low-density regions for low-to-intermediate mass galaxies in Fig. 3.

'Ram-pressure stripping' is another environment-related phenomenon. When a low-mass galaxy gets bound to the gravitational field of a galaxy cluster and moves quickly in the ICM of the cluster, the pressure exerted from the hot ICM strips the gas content of the galaxy from it (Boselli, Fossati & Sun 2022). Darvish et al. (2018b) studies the relationship between the local environment and gas content for 708 galaxies with stellar mass $\log \frac{M_*}{M_\odot} > 10$ at $0.3 < z < 4.5$. The environment is parametrized as the environmental density calculated using the adaptive kernel smoothing method (Scoville et al. 2013) and the projected comoving distance to the 10th nearest neighbour to each galaxy for high and low redshift, respectively. They find no environmental dependence for gas mass fraction (defined as the ratio of the ISM gas to the ISM gas plus stellar mass) and the depletion time-scale for their massive galaxies. Previous studies show that environmental-related processes like ram-pressure stripping can strip the gas content of galaxies in dense regions (Boselli & Gavazzi 2014). However, Darvish et al. (2018b) shows that ram-pressure stripping is only strongly effective in galaxies with stellar mass $\log \frac{M_*}{M_\odot} < 9$ with weak gravitational potential well (Fillingham et al. 2016). As a result, ram-pressure stripping is ineffective in removing the gas content of massive galaxies, including those studied in the work of Darvish et al. (2018b). Furthermore, molecular hydrogen (the main fuel for star formation) is much denser, more bound to galaxies, and consequently less vulnerable to be stripped from galaxies than atomic hydrogen. Kenney & Young (1989) and Koyama et al. (2017) find no environmental dependence for molecular hydrogen while Fumagalli et al. (2009) finds a deficiency in molecular gas in denser environments compared to low-density environments. The stellar mass range of the galaxy sample used in Darvish et al. (2018b) is the same as the most massive galaxies we explore in SIMBA (the third row of Fig. 3). We are not able to directly compare the results of these works because of the fundamental differences between the data, measurement methods, and the different definitions of galaxy properties. However, we do see a lower gas mass fraction in the two densest regions compared to lower density regions and a small difference in SFE between the densest region and the rest of the regions. This finding is not in direct contradiction with Darvish et al. (2018b) because these dense regions of SIMBA are not explored in the observations of Darvish et al. (2018b).

Lemaux et al. (2020) investigate the reversal of the star formation–density relation using the VIMOS Ultra Deep Survey. This work uses observations of 6730 star-forming galaxies with a stellar mass range $8 < \log \frac{M_*}{M_\odot} < 12$ and spectroscopic redshifts of $2 < z_{\rm spec} < 5$ in three extragalactic fields of COSMOS, Extended Chandra Deep Field South (ECDFS), and Canada-France- Hawai'i Telescope Legacy Survey (CFHTLS-D1) to explore the relationship between the SFR of galaxies and their environmental density ($\delta_{\rm gal}^*$). They use the Voronoi Monte Carlo mapping to estimate the projected density of galaxies and the environmental density contrast. In contrast to the low-redshift Universe where SFR and $\delta_{\rm gal}^*$ are anticorrelated, they find a positive correlation between these two quantities in their high-redshift sample ($2 < z < 5$). They find that this trend is mainly driven by the high fraction of massive galaxies in dense regions that are forming stars more rapidly than lower mass galaxies. Although their galaxy sample and their density estimation method are different from ours, our results in the redshift range of $2 < z < 4$ are in qualitative agreement with their findings (the top right panel of Fig. 2). A weak but statistically significant positive correlation between SFR and $\delta_{\rm gal}^*$ is still visible after controlling for mass and redshift dependence in their sample.

Wang et al. (2018) performs a similar analysis to our work on a re-simulated data set from the THREE HUNDRED PROJECT[4] (Cui et al. 2018). This data set includes 324 re-simulated clusters and four field regions from the MULTIDARK Planck simulation (Klypin et al. 2016). Wang et al. (2018) re-simulates these clusters and fields using hydrodynamical codes GADGET-X (Murante et al. 2010; Rasia et al. 2015) and GADGET-MUSIC (Sembolini et al. 2013). They divide their galaxies into three categories: (i) cluster galaxies are galaxies closer than $2R_{200}$ to the cluster centre which is the position of the most massive dark matter halo of galaxies in each re-simulated cluster; (ii) cluster vicinity galaxies which have a distance to the cluster centre between $2R_{200}$ and the fixed comoving radius of $10\,h^{-1}$ Mpc; (iii) field galaxies within a fixed radius of $38\,h^{-1}$ Mpc from the centre of the re-simulated field region, defined as the centre of the region at $z = 0$ and fixed for all redshifts. They define the SFR of each galaxy as the sum of SFRs of all gas cells of the galaxy. The SFR of each gas cell is derived from the Springel et al. (2005) prescription. They define the environment of galaxies based on the density of all matter (dark matter, stars, and gas) calculated in a $1\,h^{-1}$ Mpc sphere around each galaxy and the density contrast ($\delta_1$) is calculated compared to the average density of the universe. Their sSFR threshold to define star-forming galaxies is sSFR $> 0.3/t_{{\rm H}(z)}$. Wang et al. (2018) find that for $z = 0$ star-forming galaxies, sSFR declines when $\delta_1$ increases in all of their environment categories. However, they claim this trend is driven by the high abundance of massive galaxies in dense regions. Since massive galaxies have lower sSFRs compared to low-mass galaxies, on average, they reduce the overall sSFR in dense regions. Controlling for this effect in the data, they find that the environment does not affect galaxies in cluster and cluster vicinity, but for galaxies in the field, sSFR slightly decreases when environmental density increases. We cannot directly compare these trends to our trends for SIMBA, because at $z = 0$ many galaxies in SIMBA are not star-forming based on the definition of star-forming galaxies in Wang et al. (2018). They also show that the median sSFR for all galaxies at the redshift range of $0 < z < 2.5$ falls when $\delta_1$ increases and argue that this trend is also due to the high abundance of massive galaxies in dense regions (figure 6 of Wang et al. 2018). The difference between the sSFR-$\delta_1$ curves for their three environment categories (Cluster, vicinity, field) at low redshifts ($z = 0$) is much larger than those at high redshifts ($z > 1$), showing that the environment matters more at lower redshifts. Although the simulation configuration and measurement methods of their work are different from our work on SIMBA, a qualitative comparison between the trends shows that these results are broadly consistent. However, Wang et al. (2018) has not investigated the molecular gas content and SFE, so we cannot discuss these parameters in their simulation.

Lovell et al. (2021) introduces the First Light And Reionization Epoch Simulations (FLARES) zoom simulations and investigates the effects of the environment on the galactic properties at the epoch of reionization ($5 < z < 10$) in this simulation. Their results show a strong density dependence for the galaxy stellar mass functions and star formation rate distribution function. The galaxy star-forming main sequence, though, does not show any environmental dependence. They re-simulate spherical regions taken from a parent simulation with a box length of 3.2 comoving Gpc (Barnes et al. 2017). Using the nearest grid point mass assignment scheme, the environmental density is calculated on a $\sim 2.67$ comoving Mpc cubic grid. Then they find the environmental density on large scales by convolving this grid with a $14\,h^{-1}$ cMpc top-hat filter and define the environmental density as $\delta(x) = \rho(x)/\bar{\rho} - 1$ where $\rho(x)$ is the dark matter density on grid points and $\bar{\rho}$ is the mean density in the box. They choose 40 regions in a way that the sample includes a number of the highest overdensities (that include the first massive galaxies) and a range of different overdensities to examine the environmental dependence of galaxy formation. The redshift ranges of the FLARES simulations investigated in Lovell et al. (2021) and SIMBA studied in our work do not overlap. Moreover, the galaxy sample used in Lovell et al. (2021) contains galaxies with stellar mass in the range of $7.5 < \log \frac{M_*}{M_\odot} < 11.3$ and $7.5 < \log \frac{M_*}{M_\odot} < 10.2$ at redshifts $z = 5$ and $z = 10$, respectively. On average, they are looking at lower-mass galaxies than our work. Furthermore, FLARES includes a larger volume than SIMBA with more abundant rare and extreme environments.

### 5.3 Caveats and future work

We are aware that several biases and limitations exist in this study. For instance, hydrodynamic simulations use different numerical prescriptions to assign galaxy properties including SFR that do not completely capture the complex physics behind the formation of galaxies. Moreover, the spatial resolution of simulations are limited, making it hard to capture the subgrid physics of galaxies. Furthermore, even the periodic simulation boxes lack a sufficient number of the largest structures in the real Universe. Our work also suffers from specific caveats. We mainly explore median galaxy properties which do not represent the total distribution and scatter of all galaxies and it is not sensitive to outliers, so we might have missed some subtle variations in the distribution of galaxies. Further work is needed to check the reported trends in our work using other hydrodynamic simulations as well and confirm the physical origin of these trends.

In this work, we study the star formation activity of a large set of galaxies in a broad range of stellar mass, redshift, and environmental density. The results of this study can guide the observers to look at the galaxy sets, with specific properties, in which the environmental effects can be detected in future observational surveys. We report the most significant environmental dependences for intermediate-mass galaxies ($\log \frac{M_*}{M_\odot} < 10$) at $0 < z < 2$. Hence, observational tests of our detected trends need an estimate of the stellar mass of galaxies up to $z = 2$ that would be feasible using the large galaxy surveys to be carried out with the *James Webb Space Telescope* (*JWST*) and the Euclid mission. It is a synergy of Euclid and *JWST* that will allow studying the environmental impact on physical and structural galaxy parameters introduced in equation (10) up to $z \sim 2$. Using these parameters, observers can probe the predictions of this work.

---

[4]https://www.nottingham.ac.uk/astronomy/The300/index.php

## 6 SUMMARY AND CONCLUSION

In this work, we study the relationship between the star formation activity of simulated galaxies with stellar mass $\log \frac{M_*}{M_\odot} > 9$ and their large-scale environment from the cosmological simulation SIMBA in a redshift range of $0 < z < 4$. The environment of galaxies is defined as the environmental density of nearby galaxies, calculated on a $1\,h^{-1}$ Mpc grid using the CIC method. We explore the SFE, sSFR, and molecular hydrogen mass fraction of galaxies in different redshift, stellar mass, and environmental density bins. Additionally, we fit a scaling relation for the molecular hydrogen depletion time-scale as a function of redshift, specific star formation rate, stellar mass, radius, metallicity, and environmental density. Our most important findings include:

(i) Across the entire stellar mass range considered in this work, galaxies residing in denser regions at high redshift ($z > 1$), tend to have higher SFE than those in low-density regions. The difference is found to be $\sim 0.3$ dex at $z = 4$. This trend reverses around $z \sim 1$: at later cosmic times galaxies in dense regions have lower SFE than those in underdense regions, on average ($\sim 0.3$ dex at $z = 0$).

(ii) With a similar trend to SFE, our galaxy sample across the entire considered stellar mass range shows a weak reversal of sSFR–density relation with a turning point around $z \sim 2.5$.

(iii) Controlling for stellar mass variation, at low redshift ($z < 1$), the low- to intermediate-mass galaxies ($9 < \log \frac{M_*}{M_\odot} < 10$) in dense regions have lower SFE and sSFR than galaxies in underdense regions. The molecular hydrogen mass fraction is lower in dense regions compared to low-density regions for all galaxies regardless of their stellar mass and redshift.

(iv) We provide a scaling relation to determine depletion time-scales (and therefore SFE) in galaxies at $0 < z < 4$ with $\log \frac{M_*}{M_\odot} > 9$. The proposed fitting function takes into account the environmental density parameter, which is found to be a statistically important term in the fit.

## FUNDING

DD acknowledges support from the National Science Center (NCN) starting grant SONATA (UMO-2020/39/D/ST9/00720). LG and AWSM acknowledge the support of the Natural Sciences and Engineering Research Council of Canada (NSERC) through grant reference number RGPIN-2021-03046.

## ACKNOWLEDGEMENTS

We warmly acknowledge Behnam Darvish for providing us with their data and Sara Ellison for her thoughtful comments and discussions.

## DATA AVAILABILITY

The galaxy catalogues used in this work are available at the SIMBA project repository.[5] The full tables including the environment measurements are available as supplementary material.

---

[5] http://simba.roe.ac.uk/

# SUPPORTING INFORMATION

Supplementary data are available at *MNRAS* online.

**suppl_data**

Please note: Oxford University Press is not responsible for the content or functionality of any supporting materials supplied by the authors. Any queries (other than missing material) should be directed to the corresponding author for the article.

# APPENDIX A: TABLES

Table A1 shows the binning of galaxies based on stellar mass, redshift, and overdensity, used in Fig. 3.

**Table A1.** The number of galaxies in each bin specified by stellar mass, redshift, and the stellar mass overdensity, used in Fig. 3. The complete table can be found online as supplementary material.

| $z$ | $9 < \log M_*/M_\odot < 9.5$ | | | $z$ | $9.5 < \log M_*/M_\odot < 10$ | | | $z$ | $10 < \log M_*/M_\odot$ | | |
|---|---|---|---|---|---|---|---|---|---|---|---|
| | $\log(1+\delta^*_{\rm gal})$ | med (SFE) | $N_{\rm gal}$ | | $\log(1+\delta^*_{\rm gal})$ | med (SFE) | $N_{\rm gal}$ | | $\log(1+\delta^*_{\rm gal})$ | med (SFE) | $N_{\rm gal}$ |
| 0.0 | −0.839 988 | −9.058 414 | 1433 | 0.0 | −0.839 988 | −9.141 026 | 920 | 0.0 | −0.839 988 | −9.647 629 | 186 |
| | 0.1514 | −9.025 167 | 3744 | – | 0.1514 | −9.153 225 | 3399 | | 0.1514 | −9.600 659 | 1005 |
| | 1.142 789 | −9.072 674 | 3299 | – | 1.142 789 | −9.173 065 | 5750 | | 1.142 789 | −9.628 384 | 3164 |
| | 2.134 177 | −9.183 332 | 880 | – | 2.134 177 | −9.299 973 | 1425 | | 2.134 177 | −9.575 928 | 2458 |
| | 3.125 565 | −9.295 277 | 49 | – | 3.125 565 | −9.344 831 | 60 | | 3.125 565 | −9.430 708 | 129 |
| – | – | – | – | – | – | – | – | – | – | – | – |
| 4.0 | −0.839 988 | −8.843 436 | 191 | 4.0 | −0.839 988 | −8.690 341 | 22 | 4.0 | −0.839 988 | −8.572 915 | 4 |
| | 0.1514 | −8.830 156 | 756 | – | 0.1514 | −8.683 189 | 80 | | 0.1514 | −8.324 148 | 18 |
| | 1.142 789 | −8.836 809 | 1946 | – | 1.142 789 | −8.661 892 | 342 | | 1.142 789 | −8.351 521 | 44 |
| | 2.134 177 | −8.828 661 | 1058 | – | 2.134 177 | −8.630 128 | 521 | | 2.134 177 | −8.359 869 | 150 |
| | 3.125 565 | −8.794 724 | 168 | – | 3.125 565 | −8.605 515 | 81 | | 3.125 565 | −8.229 014 | 156 |

This paper has been typeset from a T<sub>E</sub>X/LAT<sub>E</sub>X file prepared by the author.